\pdfoutput=1

\documentclass[aps,twocolumn,superscriptaddress,floatfix,preprintnumbers,nofootinbib]{revtex4-2}

\usepackage{amsmath, float, graphicx,amssymb,multirow,pgfplots,pgfplotstable}
\usepackage{tikz,hyperref}
\usepackage{ulem}
\usepackage{comment}
\usepackage{url}

\usetikzlibrary{shapes.geometric,arrows}

\begin{document}

\begin{flushright}
YHEP-COS-21-01\end{flushright}

\title{Cosmic-Neutrino-Boosted Dark Matter ($\nu$BDM)}

\author{Yongsoo Jho}
\email{jys34@yonsei.ac.kr}
\affiliation{Department of Physics and IPAP, Yonsei University, Seoul 03722, Republic of Korea}
\author{Jong-Chul Park}
\email{jcpark@cnu.ac.kr}
\thanks{co-corresponding author}
\affiliation{Department of Physics and Institute of Quantum Systems (IQS), Chungnam National University, Daejeon 34134, Republic of Korea}
\author{Seong Chan Park}
\email{sc.park@yonsei.ac.kr}
\thanks{co-corresponding author}
\affiliation{Department of Physics and IPAP, Yonsei University, Seoul 03722, Republic of Korea}
\author{Po-Yan Tseng}
\email{tpoyan1209@gmail.com}
\affiliation{Department of Physics and IPAP, Yonsei University, Seoul 03722, Republic of Korea}

%\today

\begin{abstract}
A novel mechanism of boosting dark matter by cosmic neutrinos is proposed.
The new mechanism  is so significant that the arriving flux of dark matter in the mass window $1~{\rm keV} \lesssim m_{\rm DM} \lesssim 1~{\rm MeV}$ on Earth can be enhanced by two to four orders of magnitude compared to one only by cosmic electrons.
Thereby we firstly derive conservative but still stringent bounds and future sensitivity limits for such cosmic-neutrino-boosted dark matter ($\nu$BDM) from advanced underground experiments such as Borexino, PandaX, XENON1T, and JUNO.
\end{abstract}

\maketitle

%========================================================================
%          KEYSROKE-SAVING MACROS, nothing complicated
%========================================================================

\newcommand{\PRE}[1]{{#1}} % Use if preprint style
\newcommand{\ul}{\underline}
\newcommand{\del}{\partial}
\newcommand{\nbox}{{\,\lower0.9pt\vbox{\hrule \hbox{\vrule height 0.2 cm
\hskip 0.2 cm \vrule height 0.2 cm}\hrule}\,}}

\newcommand{\postscript}[2]{\setlength{\epsfxsize}{#2\hsize}
   \centerline{\epsfbox{#1}}}
\newcommand{\gweak}{g_{\text{weak}}}
\newcommand{\mweak}{m_{\text{weak}}}
\newcommand{\mplanck}{M_{\text{Pl}}}
\newcommand{\mstar}{M_{*}}
\newcommand{\sigmaan}{\sigma_{\text{an}}}
\newcommand{\sigmatot}{\sigma_{\text{tot}}}
\newcommand{\sigmaSI}{\sigma_{\rm SI}}
\newcommand{\sigmaSD}{\sigma_{\rm SD}}
\newcommand{\OmegaM}{\Omega_{\text{M}}}
\newcommand{\OmegaDM}{\Omega_{\text{DM}}}
\newcommand{\ipb}{\text{pb}^{-1}}
\newcommand{\ifb}{\text{fb}^{-1}}
\newcommand{\iab}{\text{ab}^{-1}}
\newcommand{\ev}{\text{eV}}
\newcommand{\kev}{\text{keV}}
\newcommand{\mev}{\text{MeV}}
\newcommand{\gev}{\text{GeV}}
\newcommand{\tev}{\text{TeV}}
\newcommand{\pb}{\text{pb}}
\newcommand{\mb}{\text{mb}}
\newcommand{\cm}{\text{cm}}
\newcommand{\m}{\text{m}}
\newcommand{\km}{\text{km}}
\newcommand{\kg}{\text{kg}}
\newcommand{\g}{\text{g}}
\newcommand{\s}{\text{s}}
\newcommand{\yr}{\text{yr}}
\newcommand{\Mpc}{\text{Mpc}}
\newcommand{\etal}{{\em et al.}}
\newcommand{\eg}{{\em e.g.}}
\newcommand{\ie}{{\em i.e.}}
\newcommand{\ibid}{{\em ibid.}}
\newcommand{\Eqref}[1]{Equation~(\ref{#1})}
\newcommand{\secref}[1]{Sec.~\ref{sec:#1}}
\newcommand{\secsref}[2]{Secs.~\ref{sec:#1} and \ref{sec:#2}}
\newcommand{\Secref}[1]{Section~\ref{sec:#1}}
\newcommand{\appref}[1]{App.~\ref{sec:#1}}
\newcommand{\figref}[1]{Fig.~\ref{fig:#1}}
\newcommand{\figsref}[2]{Figs.~\ref{fig:#1} and \ref{fig:#2}}
\newcommand{\Figref}[1]{Figure~\ref{fig:#1}}
\newcommand{\tableref}[1]{Table~\ref{table:#1}}
\newcommand{\tablesref}[2]{Tables~\ref{table:#1} and \ref{table:#2}}
\newcommand{\Dsle}[1]{\slash\hskip -0.28 cm #1}
\newcommand{\met}{{\Dsle E_T}}
\newcommand{\mpt}{\not{\! p_T}}
\newcommand{\Dslp}[1]{\slash\hskip -0.23 cm #1}
\newcommand{\Dsl}[1]{\slash\hskip -0.20 cm #1}

\newcommand{\mB}{m_{B^1}}
\newcommand{\mq}{m_{q^1}}
\newcommand{\mf}{m_{f^1}}
\newcommand{\mKK}{m_{KK}}
\newcommand{\WIMP}{\text{WIMP}}
\newcommand{\SWIMP}{\text{SWIMP}}
\newcommand{\NLSP}{\text{NLSP}}
\newcommand{\LSP}{\text{LSP}}
\newcommand{\mWIMP}{m_{\WIMP}}
\newcommand{\mSWIMP}{m_{\SWIMP}}
\newcommand{\mNLSP}{m_{\NLSP}}
\newcommand{\mchi}{m_{\chi}}
\newcommand{\mgravitino}{m_{\gravitino}}
\newcommand{\mmed}{M_{\text{med}}}
\newcommand{\gravitino}{\tilde{G}}
\newcommand{\Bino}{\tilde{B}}
\newcommand{\photino}{\tilde{\gamma}}
\newcommand{\stau}{\tilde{\tau}}
\newcommand{\slepton}{\tilde{l}}
\newcommand{\snu}{\tilde{\nu}}
\newcommand{\squark}{\tilde{q}}
\newcommand{\mgaugino}{M_{1/2}}
\newcommand{\epsEM}{\varepsilon_{\text{EM}}}
\newcommand{\mmess}{M_{\text{mess}}}
\newcommand{\lmess}{\Lambda}
\newcommand{\nmess}{N_{\text{m}}}
\newcommand{\signmu}{\text{sign}(\mu)}
\newcommand{\Omegachi}{\Omega_{\chi}}
\newcommand{\lambdafs}{\lambda_{\text{FS}}}
\newcommand{\be}{\begin{equation}}
\newcommand{\ee}{\end{equation}}
\newcommand{\bea}{\begin{eqnarray}}
\newcommand{\eea}{\end{eqnarray}}
\newcommand{\beq}{\begin{equation}}
\newcommand{\eeq}{\end{equation}}
\newcommand{\beqn}{\begin{eqnarray}}
\newcommand{\eeqn}{\end{eqnarray}}
\newcommand{\baln}{\begin{align}}
\newcommand{\ealn}{\end{align}}
\newcommand{\lsim}{\lower.7ex\hbox{$\;\stackrel{\textstyle<}{\sim}\;$}}
\newcommand{\gsim}{\lower.7ex\hbox{$\;\stackrel{\textstyle>}{\sim}\;$}}

\newcommand{\ssection}[1]{{\em #1.\ }}
\newcommand{\rem}[1]{\textbf{#1}}

\def\ie{{\it i.e.}\/}
\def\eg{{\it e.g.}\/}
\def\etc{{\it etc}.\/}
\def\calN{{\cal N}}

\def\mptwo{{m_{\pi^0}^2}}
\def\mp{{m_{\pi^0}}}
\def\sqtsn{\sqrt{s_n}}
\def\sqtsn{\sqrt{s_n}}
\def\sqtsn{\sqrt{s_n}}
\def\sqts0{\sqrt{s_0}}
\def\Dsqts{\Delta(\sqrt{s})}
\def\Omegatot{\Omega_{\mathrm{tot}}}

\newcommand{\changed}[2]{{\protect\color{red}\sout{#1}}{\protect\color{blue}\uwave{#2}}}

%%%%%%%%%%%%%%%%%%%%%%%%%%%%%%%%%%%%%%%%%%%%%%%%%%%%%%%%%%%%%%%%%%%%%%%%%%%%%%%

\section{Introduction}

Revealing the properties of dark matter (DM) is definitely one of the most pressing issues in particle physics, astrophysics, and cosmology.
Direct detection experiments of DM have particular importance as they aim to probe  interaction of DM with standard model (SM) particles~\cite{Goodman:1984dc}.
However, there exists fundamental limitation in detecting a sub-MeV dark matter set by the maximum kinetic energy of the DM particle in halo:
\begin{equation}
K_{\rm DM}^{\rm max} \lsim 10^{-6} m_{\rm DM} \lsim 1 ~{\rm eV}
\end{equation}
with the velocity $v\sim 10^{-3}$.
This low kinetic energy causes a significant problem in detecting light dark matter since the recoil energy of scattered SM particle is also limited by the kinetic energy~\footnote{Several ideas have been suggested to detect signals with low recoil energies by lowering the threshold energies at detectors (see \cite{Battaglieri:2017aum, Kim:2020bwm} and references therein).}.
On the other hand, there still exists a chance to detect a subcomponent of DM, dubbed `boosted dark matter' (BDM), which may carry much larger energy beyond threshold due to various mechanisms~\cite{Belanger:2011ww, Agashe:2014yua, Berger:2014sqa, Kong:2014mia, Kim:2016zjx, Giudice:2017zke, DEramo:2010keq} including scattering by energetic cosmic-ray particles~\cite{Bringmann:2018cvk, Ema:2018bih,  Cappiello:2018hsu, Cappiello:2019qsw, Dent:2019krz, Cho:2020mnc, Jho:2020sku}.
We note that focus has been given to cosmic-ray electron and proton so far even though the chance is not exclusively open for charged particles.

{
In this letter, we  focus on a noble class of cosmic-neutrino-boosted-dark matter ($\nu$BDM) extending previous studies: there exist a huge number of cosmic-ray neutrinos arriving at the solar system from various origins~\cite{Vitagliano:2019yzm}.
Our Sun is also generating a large number of neutrinos~\cite{Bahcall:2004pz, Billard:2013qya, Vitagliano:2017odj} so that they may boost DM within the solar system.
We find that $\nu$BDM can be a dominant part of the whole BDM when DM-neutrino interaction is as strong as DM-electron interaction, which is indeed the case for gauged lepton number as mediator, for instance \cite{Rajpoot:1989jb, He:1990pn}.
The existing conclusions regarding cosmic-electron-induced BDM should be re-examined.  }

\section{Boost mechanism by cosmic neutrino}

%- In this section, only {\it model-independent} analysis with $\sigma_{\chi i}$, $i$ being SM particles including neutrino for {\it boosting} and electron (nucleon?) for {\it detection}.

\noindent {\bf Cosmic neutrino inputs.}
%- fluxes and spectra~\cite{Billard:2013qya, Vitagliano:2017odj}, etc (for mainly %solar/star neutrinos)
Near Earth, our Sun provides the dominating neutrino flux $d\Phi^{\rm Sun}_\nu/dK_\nu$ in the neutrino energy $K_\nu \lesssim 10$~MeV reaching the maximum $\simeq \mathcal{O}(10^{8})~[{\rm cm^{-2}\, s^{-1}\, keV^{-1}}]$ around $K_\nu \simeq 0.3$ MeV~\cite{Bahcall:2004pz, Billard:2013qya, Vitagliano:2017odj}, which gives the total number of neutrino emission rate per unit energy
\begin{eqnarray}
\frac{d\dot{N}^{\rm Sun}_\nu}{dK_\nu} \equiv
\frac{d\Phi^{\rm Sun}_\nu}{dK_\nu}\,(4\pi D^2_\odot)\,,
\end{eqnarray}
where $D_\odot=1$~AU is the distance between Sun and Earth.
The neutrinos can boost non-relativistic light DM, leaving distinctive signals at terrestrial experiments, e.g. XENON1T~\cite{Aprile:2017aty, Aprile:2020tmw}.
The total contributions from all stars for $\nu$BDM could be significant compared to the BDM flux by the solar neutrinos.
The overall neutrino flux from all stars in the Milky Way (say, {\it cosmic-neutrino} flux) has not been measured by astrophysical observations, and could be highly anisotropic, which is different from the isotropic diffused cosmic electrons.
In general, DM particles can be boosted by the neutrino flux from the nearest star, instead of diffused neutrinos.
Keep this philosophy in mind, we will compute the $\nu$BDM flux by starting with single star contribution in the following section, then integrate the entire star distribution in the Milky Way.

\medskip
\noindent {\bf Cosmic neutrino and DM scattering.}
The halo DM is boosted by neutrino through the process $\nu+\chi \to \nu+\chi$, which may originate from the exchange of the $U(1)_{L_e-L_i}$ gauge boson or dim-6 effective operators including $(\bar{\ell}\gamma^\mu \ell)(\bar{\chi}\gamma_\mu \chi)$ or $(\bar{\ell} \ell)(\bar{\chi} \chi)$.
The resulting BDM kinetic energy $K_{\rm DM}$ can be determined from the kinetic energy of incoming neutrino $K_\nu$.
At the halo DM rest frame, the allowed range of $K_{\rm DM}$ is given by~\cite{Boschini:2018zdv}
\begin{eqnarray}
0 \leq K_{\rm DM} \leq K^{\rm max}_{\rm DM} \equiv \frac{2m_{\rm DM} (K^2_\nu+2m_\nu K_\nu)}{(m_{\rm DM}+m_\nu)^2+2m_{\rm DM}K_\nu}\,.
\end{eqnarray}

\begin{figure}[t]
\centering
\includegraphics[width=0.40\textwidth]{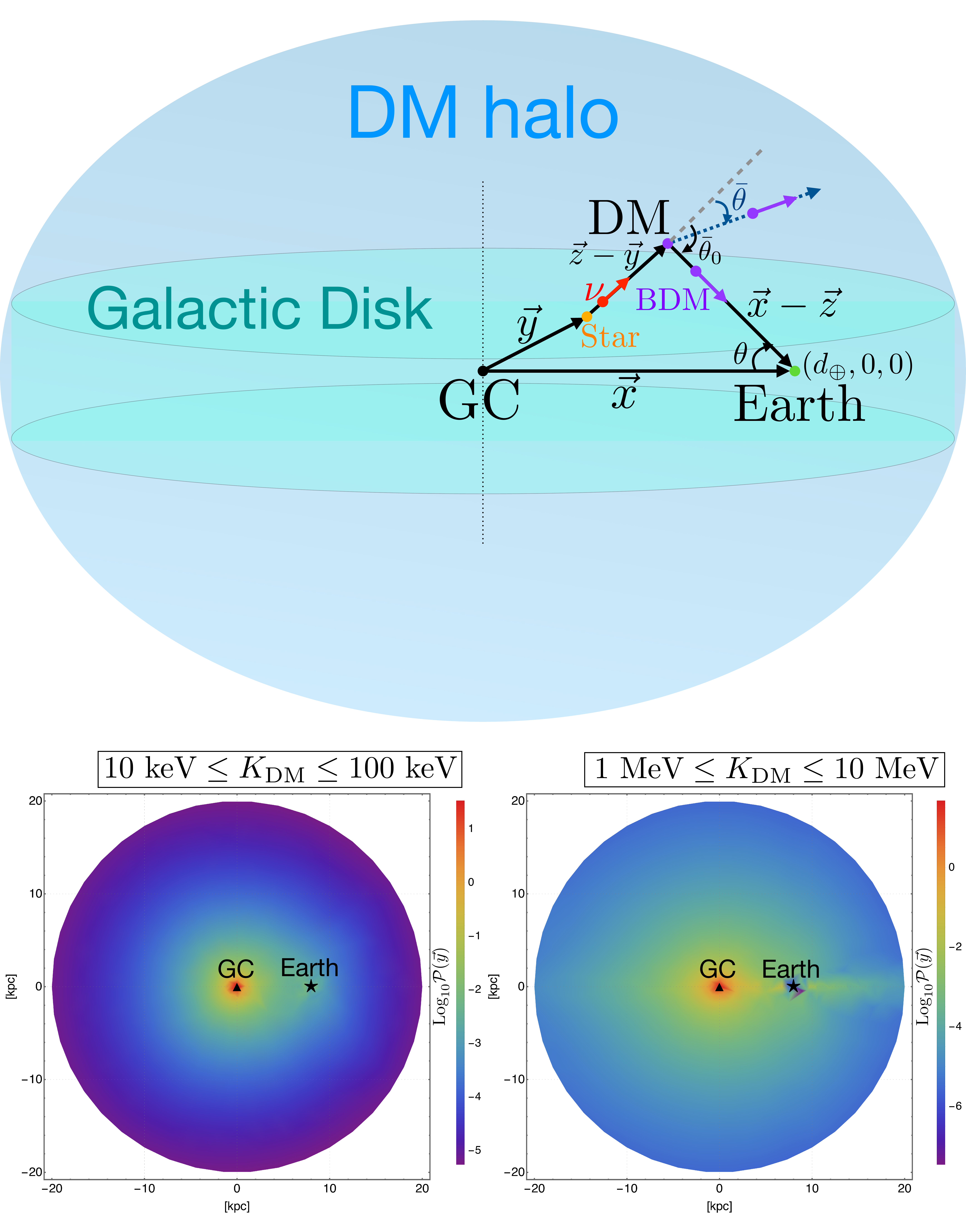}
\caption{[{\bf Top}] Schematic description of BDM production by the neutrino from a single star.
[{\bf Bottom}] Areal density of unit-normalized distribution of the $\nu$BDM flux from stars in our galaxy $\mathcal{P}(\vec{y}) \equiv \frac{1}{\Phi_{\rm DM}} \frac{d\Phi_{\rm DM}}{dA_y}$ [kpc$^{-2}$], for two representative ranges of $K_{\rm DM}$: 10 -- 100 keV (left) and 1 -- 10 MeV (right).  $dA_y$ is the areal element of the Galactic disk, defined by position of star, $\vec{y}$.  \label{fig:schematic}
}
\end{figure}

The BDM flux by neutrinos from a Sun-like star is
\begin{eqnarray}
\label{eq:DMflux}
\frac{d\Phi^{(1)}_{\rm DM}(\overrightarrow{y})}{dK_{\rm DM}}
&\simeq &
\frac{1}{8\pi^2}
\left( \tilde{f}_1 \frac{d\dot{N}^{\rm Sun}_\nu}{dK_\nu} \right)
\int d^3 \overrightarrow{z}
\frac{\rho_{\rm DM}(|\overrightarrow{z}|)}{m_{\rm DM}}
\frac{1}{|\overrightarrow{x}-\overrightarrow{z}|^2} \nonumber \\
&& \times \left( \left.
\frac{dK_\nu}{d\bar{\theta}}\right\vert_{\bar{\theta}=\bar{\theta}_0}
 \right)
 \left(
 \left. \frac{d\sigma_{\nu{\rm DM}}}{dK_{\rm DM}}\right\vert_{\bar{\theta}=\bar{\theta}_0}
 \right) \nonumber \\
&&\times
\frac{1}{\sin\bar{\theta}_0}\frac{1}{|\overrightarrow{z}-\overrightarrow{y}|^2}
\times \exp {\left(-\frac{|\overrightarrow{z}-\overrightarrow{y}|}{d_\nu} \right)} \,,
\end{eqnarray}
where the schematic diagram of the coordinate system is shown in the top panel of Fig.~\ref{fig:schematic}, and $\overrightarrow{x},\overrightarrow{y}$, and $\overrightarrow{z}$ represent the positions of Earth, Star, and halo DM, respectively.
The correction factor $\tilde{f}_1$ takes into account the variances of stellar properties from Sun~\cite{Farag:2020nll} and $\rho_{\rm DM}$ is the DM halo density profile.
The differential $\nu$-DM cross section depends on scattering angle $\bar{\theta}$, and $\bar{\theta}_0$ can be determined by $K_\nu$ and $K_{\rm DM}$ via kinematic relations:
\begin{eqnarray}
\label{eq:kinematic}
K_\nu(K_{\rm DM},\bar{\theta})&=& \frac{p'^2-K^2_{\rm DM}}
{2\left(p'\cos \bar{\theta}-K_{\rm DM} \right)}\,, \\
\left.\frac{dK_\nu}{d\bar{\theta}}\right\vert_{\bar{\theta}=\bar{\theta}_0}&=&
\frac{(p'^2-K^2_{\rm DM})p'}
{2\left(p'\cos \bar{\theta}_0-K_{\rm DM} \right)^2}
\sin \bar{\theta}_0\,,
\end{eqnarray}
where $p'\equiv \sqrt{2m_{\rm DM}K_{\rm DM}+K^2_{\rm DM}}$ is 3-momentum of BDM in the halo DM frame.
$dK_{\nu}/d\bar{\theta}\propto 1/\cos^{2}\bar{\theta}$ and large scattering angle $\bar{\theta}\simeq \pi/2$ is favoured for $m_{\rm DM} \gg K_{\rm DM}$, whereas  $dK_{\nu}/d\bar{\theta}\propto 1/(\cos\bar{\theta}-1)^2$ makes the forward scattering $\bar{\theta}\simeq 0$ dominate for $m_{\rm DM} \ll K_{\rm DM}$.

The neutrino flux attenuation due to propagation is determined by the exponential function in Eq.~(\ref{eq:DMflux}), and the mean free path is obtained as $d_\nu \equiv 1/[(\rho_{\rm DM}/m_{\rm DM})\cdot \sigma_{\nu{\rm DM}}]$ where the total $\nu$-DM cross section is
\begin{eqnarray}
\sigma_{\nu{\rm DM}}(K_\nu) \equiv \int^{K^{\rm max}_{\rm DM}}_0 dK_{\rm DM} \frac{d \sigma_{\nu{\rm DM}}}{dK_{\rm DM}}\,.
\end{eqnarray}

In the derivation of Eq.~(\ref{eq:DMflux}), we use point-like star approximation,
by starting with finite star radius $R_{\rm star}$ then taking $R_{\rm star} \to 0$.
The final result of $d\Phi^{(1)}_{\rm DM}(\overrightarrow{y})/dK_{\rm DM}$ is finite.
Due to the distance-squared suppression, the dominating $\nu$BDM fluxes originate either from halo DM at the vicinity of Earth or the galatic center (GC).

\begin{figure}[t]
\centering
{\includegraphics[width=0.45\textwidth]{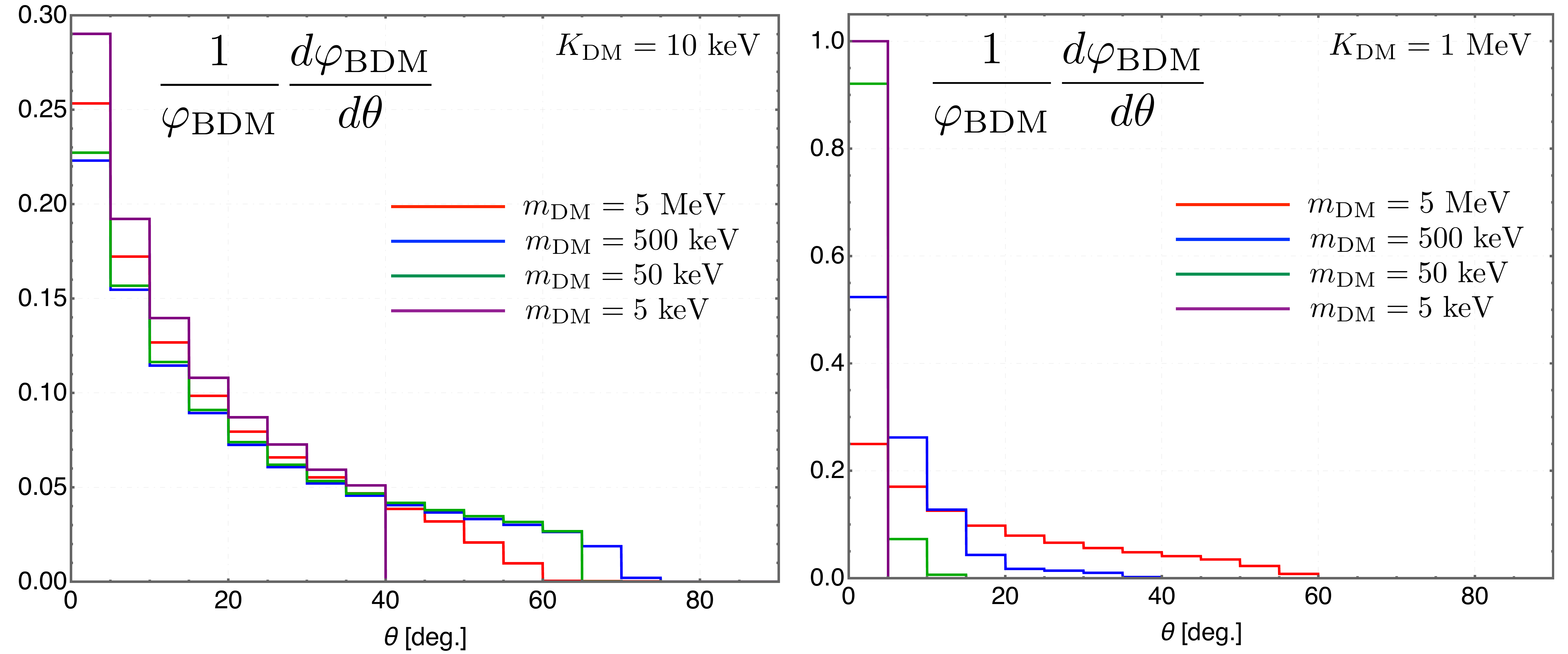}}
\caption{The unit-normalized arrival direction $\theta$ distributions of the $\nu$BDM spectral flux $\varphi_{\rm BDM} \equiv d\Phi_{\rm DM}/dK_{\rm DM}$ for two benchmark values of $K_{\rm DM}$: 10 keV (left) and 1 MeV (right) varying $m_{\rm DM}=$ 5 keV -- 5 MeV with a fixed mediator mass, $m_X = 700$ keV. \label{fig:BDMflux_dist}}
\end{figure}

From Eq.~(\ref{eq:DMflux}), we can calculate the BDM flux by neutrinos from Sun by taking $|\overrightarrow{x}-\overrightarrow{y}|=D_\odot$.
Even though Sun provides the largest neutrino flux to Earth, only small volume of nearby DM halo compromises the BDM flux.
Therefore, we need to consider the entire stellar contributions in the Milky Way by convolving Eq.~(\ref{eq:DMflux}) with stellar distribution $n_{\rm star}(\overrightarrow{y})$:
\begin{eqnarray} \label{eq:DMflux_total}
\frac{d\Phi_{\rm DM}}{dK_{\rm DM}} = \int d^3 \overrightarrow{y} n_{\rm star}(\overrightarrow{y}) \frac{d\Phi^{(1)}_{\rm DM}(\overrightarrow{y})}{dK_{\rm DM}}\,.
\end{eqnarray}
Here we assume stars distribute within the galactic disk, shown in the top panel of Fig.~\ref{fig:schematic}, with radius $R\leq 20$ kpc and thickness $|h|\leq 1$ kpc.
Using the observation~\cite{deJong:2009iq} and integrating out the $h$, the stellar distribution on 2-dimensional galactic disk is given by
\begin{eqnarray} \label{eq:star_distribution}
n_{\rm star}(R) \simeq \tilde{f}_2 \times 1.2\times 10^{11} / (R/{\rm kpc})^3~[{\rm kpc^{-2}}]\,,
\end{eqnarray}
where $\tilde{f}_2$ factor includes the uncertainties from detailed structures of the Milky Way, e.g. spiral arms and density fluctuations.

The $\nu$BDM fluxes with two $K_{\rm DM}$ regimes are shown in the bottom panels of Fig.~\ref{fig:schematic}.
Due to the high stellar and DM number densities around the GC, the BDM flux contribution from the GC region exceeds that from the vicinity of Earth.
Fig.~\ref{fig:BDMflux_dist} shows the $\theta$ dependence of
the $\nu$BDM fluxes $\frac{1}{\varphi_{\rm BDM}} \frac{\varphi_{\rm BDM}}{d\theta}$ at Earth where $\theta$ represents the angle between the $\nu$BDM arrival direction and the GC.
For $K_{\rm DM} \gg m_{\rm DM}$ in the right panel, the forward scattering ($\bar{\theta}_0\simeq 0$) is preferred, so that $\nu$BDM from the GC dominates and thus $\theta\simeq 0$.
In the left panel, $K_{\rm DM} \ll m_{\rm DM}$ prefers large-angle scattering ($\bar{\theta}_0\simeq 90^\circ$), which enhances the flux for $\theta \gtrsim 40^\circ$ originating relatively far from the GC.
The $\theta$ dependence of the $\nu$BDM flux can be used to determine $m_{\rm DM}$ in the future.

\begin{figure}[t]
\centering
\includegraphics[width=7.0cm]{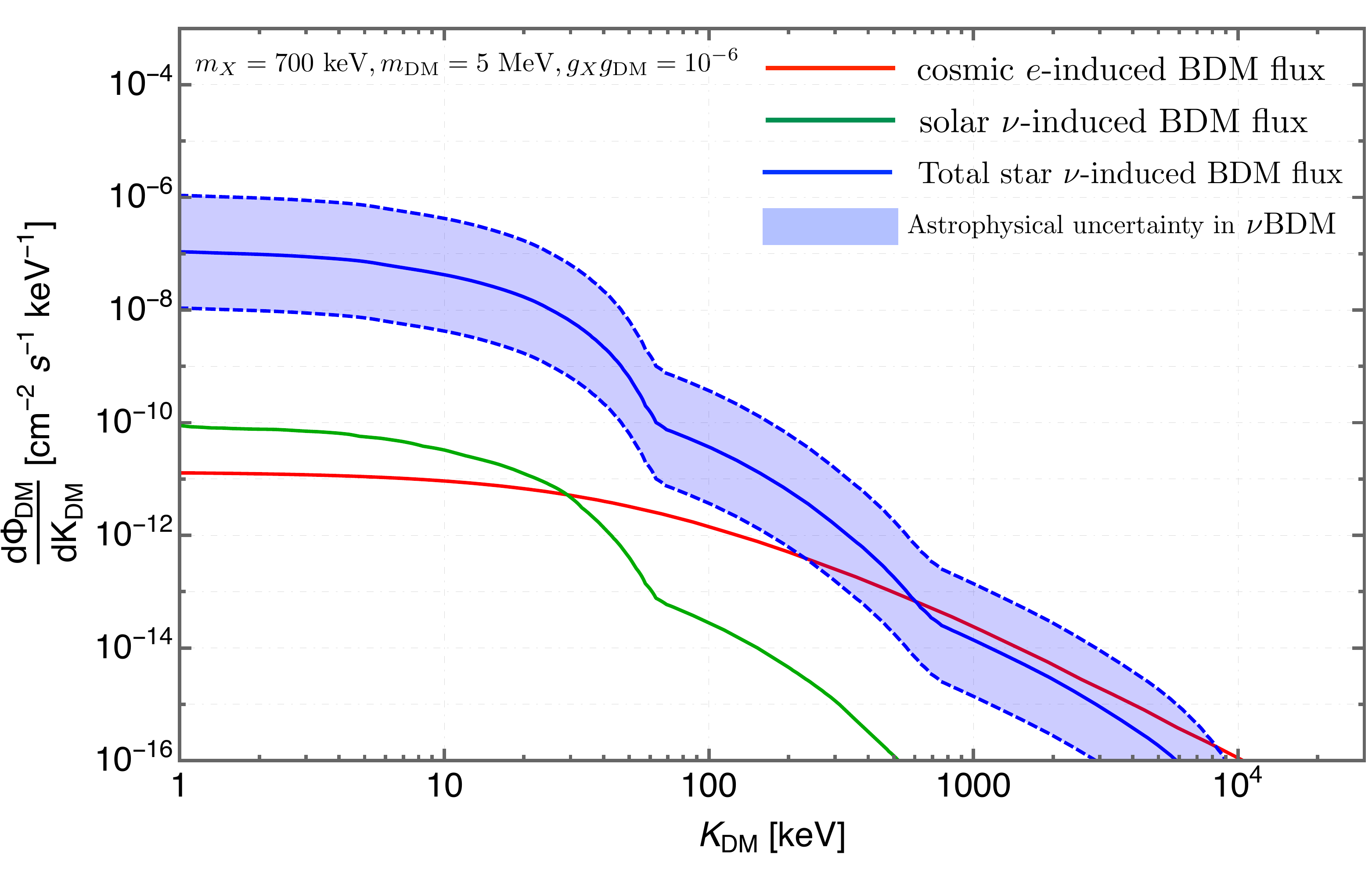}
\includegraphics[width=7.0cm]{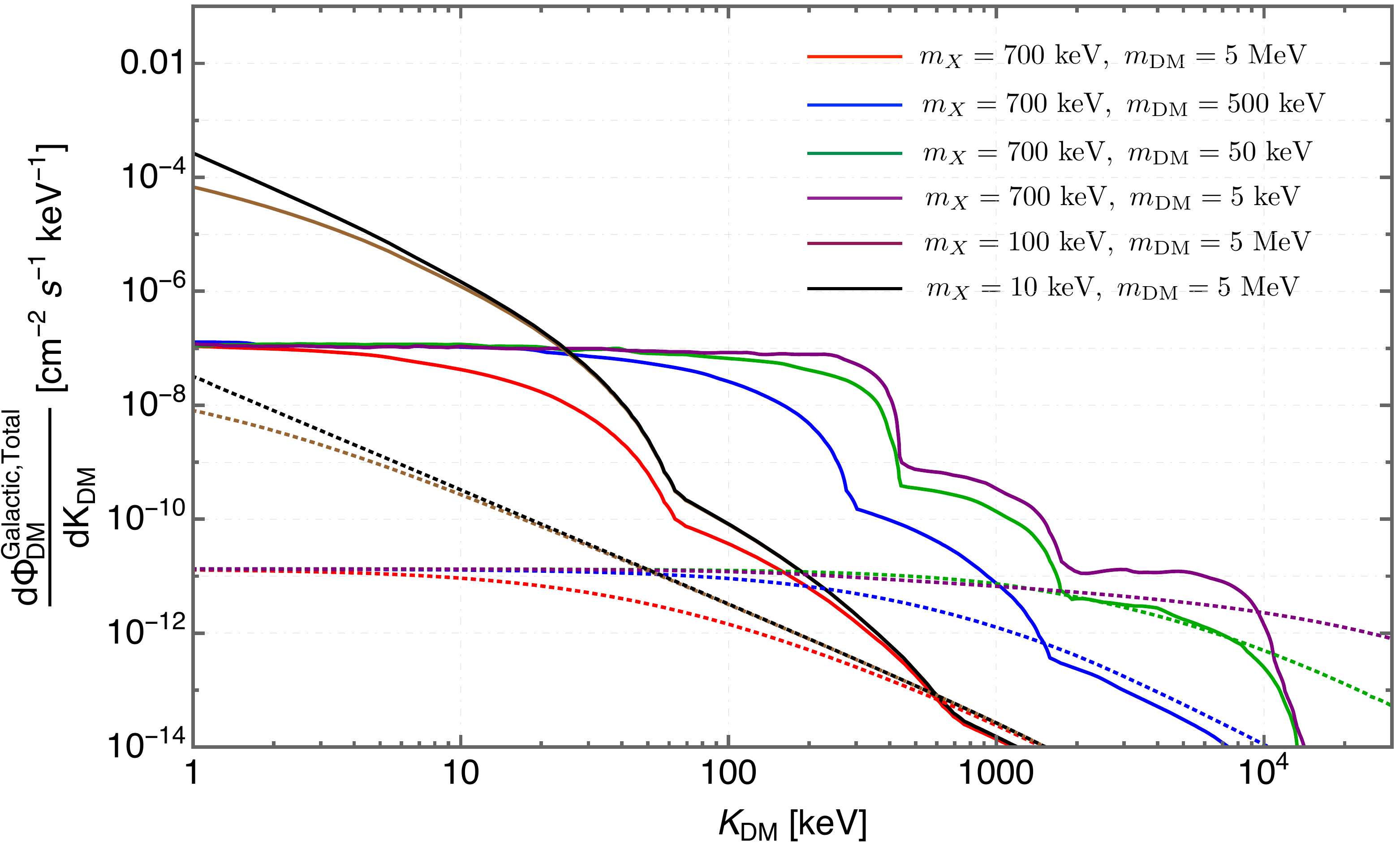}
\caption{{\bf [Top]} BDM fluxes by solar neutrinos, cosmic neutrinos, and cosmic electrons.
We assume $\sigma_{\nu{\rm DM}}$ comes from a vector boson $X$ coupling to both DM and leptons ($g_X = g_e = g_\nu$) with $(m_{\rm DM}, m_X, g_X g_{\rm DM})
=(5 {\rm MeV}, 700 {\rm keV}, 10^{-6})$.
The uncertainty band for $\nu$BDM corresponds to $0.1 \leq\tilde{f}\leq 10$.
{\bf [Bottom]} BDM fluxes for different $m_X$ and $m_{\rm DM}$ with $\tilde{f} = 1$.
Solid and dotted lines are $\nu$BDM and cosmic electron BDM fluxes, respectively. \label{fig:total_DMflux}
}
\end{figure}

In the top panel of Fig.~\ref{fig:total_DMflux}, we compare the BDM fluxes via solar neutrinos, {\it cosmic neutrinos}, and cosmic-ray electrons by fixing $\tilde{f} \equiv \tilde{f}_1 \cdot \tilde{f}_2$ =1.
The $\nu$BDM flux is three orders of magnitude larger than that by solar neutrinos, because the later is relevant to DM only within a few AUs around Earth.
Three bumps of the $\nu$BDM flux correspond to the $pp$, $^{13}$N+$^{15}$O, and $^8$B production processes of solar neutrinos~\cite{Bahcall:2004pz}.
Assuming $g_e = g_\nu$, the $\nu$BDM flux can be two to four orders of magnitude larger than that induced by cosmic electrons for $K_{\rm DM}\lesssim 50~{\rm keV}$.
This feature is quite robust for other DM and mediator masses as shown in the bottom panel.

There are several factors that can make our estimations different.
{\it i\,}) The DM halo profile, especially around the GC.
We take the NFW profile.
{\it ii\,}) The $\nu$ flux varies with the type and age of stars~\cite{Farag:2020nll}.
{\it iii\,}) The star distribution in the Milky Way.
All the above uncertainties are hard to be included in the calculation.
In order to show the robustness of the results, we conservatively vary $
0.1 \lesssim \tilde{f} \lesssim 10$ in Eq.~(\ref{eq:DMflux}) and (\ref{eq:DMflux_total}), depicted as a blue band in the top panel of Fig.~\ref{fig:total_DMflux}.

\medskip
\noindent {\bf Attenuation of the BDM flux.}
The attenuation effect of the {\it cosmic-neutrino} flux scattered by halo DM is taken into account by the exponential factor in Eq.~\ref{eq:DMflux}.
We estimate the mean free path of cosmic neutrino $d_\nu$ by taking $\sigma_{\nu{\rm DM}}\simeq 10^{-28}-10^{-34}~{\rm cm^2}$.
For the $n_{\rm DM} \sim ({\rm keV}/m_{\rm DM})\times 10^{6} ~{\rm cm^{-3}}$, $d_\nu\simeq (m_{\rm DM}/{\rm keV}) \times (10^{22}-10^{28})~{\rm cm}$, which is larger than the size of the Milky Way and results in negligible effect.

Next, we estimate the mean free path of BDM inside Earth by assuming  $\sigma_{e{\rm DM}}=\sigma_{\nu{\rm DM}}$.
For $\sigma_{e{\rm DM}}=10^{-33}~{\rm cm^2}$ with electron number density of Earth $n_e\simeq 10^{24}~{\rm cm^{-3}}$, the mean free path $1/(n_e \cdot \sigma_{e{\rm DM}})\simeq 10^4~{\rm km}$ is comparable to the size of Earth.
For $\sigma_{e{\rm DM}}=10^{-29}~{\rm cm^2}$, the BDM mean free path reduces to $\mathcal{O}({\rm km})$.
Most of DM direct detection detectors locate a few kilometers underground, rendering the $\nu$BDM signal be substantially suppressed for $\sigma_{e{\rm DM}} \gtrsim 10^{-29}~{\rm cm^2}$.
Thus, the attenuation of BDM inside Earth will provide upper limits on experimental sensitivities as shown in Fig.~\ref{fig:sigmaeff_scan}.

\section{Experimental sensitivities}

To estimate experimental sensitivities, we use two approaches for DM models.
{\it i\,}) Heavy mediator: the interactions can be described by effective cross sections $\sigma^{\rm eff}_{\nu{\rm DM}}$ and $\sigma^{\rm eff}_{e{\rm DM}}$.
{\it ii\,}) Light mediator: the $X$ boson from the gauged $U(1)_{L_e - L_i}$ couples to DM and leptons.

For the approach {\it i\,}), the differential cross section is defined as
\begin{eqnarray}
\frac{d\sigma^{\rm eff}_{\nu{\rm DM}, e{\rm DM}}}{dK_{\rm DM}} \equiv \frac{\sigma^{\rm eff}_{\nu{\rm DM}, e{\rm DM}}}{K^{\rm max}_{\rm DM}-K^{\rm min}_{\rm DM}}\,.
\end{eqnarray}
On the other hand, for the $U(1)_{L_e - L_i}$ model, the neutrino-DM scattering cross section is given by~\cite{Cao:2020bwd}
\begin{widetext}
\begin{align}
& \frac{d\sigma_{\nu{\rm DM}}}{dK_{\rm DM}} = \frac{(g_X g_{\rm DM})^2}{4\pi} \frac{2m_{\rm DM}(m_\nu+K_\nu)^2 - K_{\rm DM}\left[(m_\nu + m_{\rm DM})^2+2m_{\rm DM}K_\nu\right] + m_{\rm DM}K^2_{\rm DM}}{(2m_\nu K_\nu+K^2_\nu)(2m_{\rm DM} K_{\rm DM}+m^2_X)^2} \,.
\end{align}
\end{widetext}
For $K_{\rm DM}\simeq \mathcal{O}({\rm keV})$ and $m_{\rm DM}\simeq m_X \simeq \mathcal{O}({\rm MeV})$, it makes $d\sigma_{\nu{\rm DM}}/dK_{\rm DM}$ almost independent of $K_{\rm DM}$.

\begin{figure}[t]
\centering
\includegraphics[height=7.cm,angle=270]{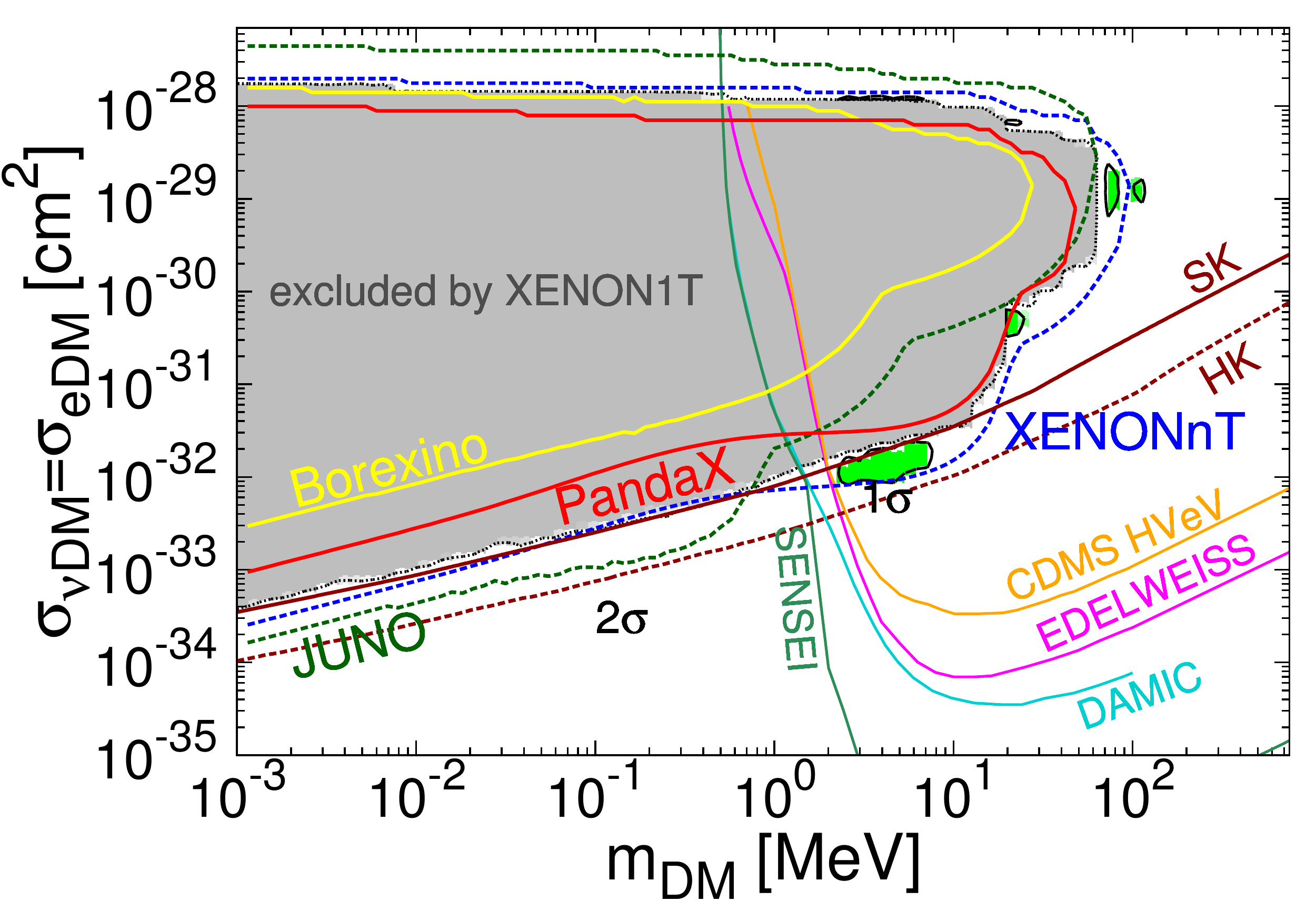}
\caption{\small \label{fig:sigmaeff_scan}
$\nu$BDM contributions to XENON1T electron recoil, assuming $\sigma^{\rm eff}_{\nu{\rm DM}}=\sigma^{\rm eff}_{e{\rm DM}}$, where the $1\sigma$ (green) and $2\sigma$ (white) regions from $\chi^2$ analysis, and the gray-shaded region is excluded more than $2\sigma$.
The expected sensitivities from other underground detectors are depicted: Brexino~\cite{Bellini:2011rx}, PandaX~\cite{Zhou:2020bvf}, XENONnT~\cite{Aprile:2020vtw}, and JUNO~\cite{An:2015jdp}.
For comparison, existing limits are shown together: CDMS HVeV~\cite{Agnese:2018col}, DAMIC~\cite{Aguilar-Arevalo:2019wdi}, EDELWEISS~\cite{Arnaud:2020svb}, and SENSEI~\cite{Barak:2020fql}.
The cosmic-electron-BDM constraints from Super-K and Hyper-K~\cite{Cappiello:2019qsw}.
}
\end{figure}

We perform the model-independent $\chi^2$ analysis for the effective cross section $(m_{\rm DM}, \sigma^{\rm eff}_{\nu{\rm DM}}=\sigma^{\rm eff}_{e{\rm DM}})$ in Fig.~\ref{fig:sigmaeff_scan}.
There are five disconnected $1\sigma$ regions for the XENON1T excess~\cite{Aprile:2020tmw}, which originate from the three bumps of the $\nu$BDM flux spectrum in Fig.~\ref{fig:total_DMflux}.
The $2\sigma$ exclusion region is gray-shaded.
The $\nu$BDM provides stringent constraint on $\sigma^{\rm eff}_{\nu{\rm DM}}=\sigma^{\rm eff}_{e{\rm DM}}$ for unexplored small mass $m_{\rm DM}\lesssim {\rm MeV}$, compared with the current limits from DM direct detection experiments including CDMS HVeV~\cite{Agnese:2018col}, DAMIC~\cite{Aguilar-Arevalo:2019wdi}, EDELWEISS~\cite{Arnaud:2020svb}, and SENSEI~\cite{Barak:2020fql}.

We evaluate the sensitivities of $\nu$BDM with other current (Brexino~\cite{Bellini:2011rx}, PandaX~\cite{Zhou:2020bvf}) and future experiments (XENONnT~\cite{Aprile:2020vtw}, JUNO~\cite{An:2015jdp}).
To estimate the sensitivities, we take the four ton-year exposure for XENONnT and 20 kton-year exposure for JUNO assuming no excess above the expected background and dominance of statistical uncertainty.
Borexino and JUNO have higher energy threshold above 100 keV but huge statistics.
JUNO has the best sensitivity for $m_{\rm DM}\lesssim 0.5~{\rm MeV}$, while XENON1T/nT are better than JUNO for $m_{\rm DM}\gtrsim 0.5~{\rm MeV}$.
PandaX has a slightly weaker limit due to the smaller 0.276 ton-year exposure than XENON1T of 0.65 tonne-year.
For $\sigma_{e{\rm DM}}\gtrsim 10^{-29}~{\rm cm^2}$, the earth crust attenuates the BDM flux; specifically, XENON1T and XENONnT~\cite{Aprile:2017aty} are located underground at a depth of 3600 meter water equivalent (m.w.e.) and Borexino~\cite{Back:2012awa} is at 3800 m.w.e while PandaX~\cite{Cao:2014jsa} is shielded by 2400 m marble overburden ($\sim6800$ m.w.e.).
The most shallow JUNO detector~\cite{An:2015jdp}, located at 700 m deep underground ($\sim2000$ m.w.e.), has the best upper sensitive to $\nu$BDM with $\sigma_{e{\rm DM}} \simeq 10^{-28}~{\rm cm^2}$.

\vspace{.5cm}

\section{Discussions}
The flux of the cosmic-neutrino-boosted-DM ($\nu$BDM) is substantially larger than the one of the cosmic-electron-boosted-DM so that it contributes dominantly in direct detection experiments on Earth.  Due to the distributions of the sources of neutrinos in Milky Way and the dark matter in halo, the angular distribution of the $\nu$BDM is kinematically correlated with the DM mass. Therefore precise measurement of directional information helps in determination of the DM mass.

The existing underground detectors probe the parameter region of neutrino-DM interaction and electron-DM interaction in $10^{-34}~{\rm cm^2}\lesssim \sigma_{\nu{\rm DM}}=\sigma_{e{\rm DM}}\lesssim 10^{-28}~{\rm cm^2}$
with $1~{\rm keV} \lesssim m_{\rm DM} \lesssim 100~{\rm MeV}$ based on the effective cross section approach.
Since the DM flux is enhanced by neutrino-boost, we find parameter regions for the recent XENON1T anomaly (see Fig.~\ref{fig:sigmaeff_scan}). However, they are still hardly consistent with other DM searches.

Finally, we discuss various factors of future refinement of the current study. Here we only assumed that nuclear activities inside each star are on average same as in our Sun,  so that the neutrino fluxes from each star are all similar.  Obviously, this is a crude estimation and actual neutrino fluxes differ from star to star. Also, the GC region has the largest population of main sequence stars and also red giants~\cite{Robin:2011fw, Bulge1}, which enhances $ \tilde{f}_1 \cdot \tilde{f}_2$ factor over unity~\cite{Farag:2020nll}. Last but not least, we point out the potential modification due to the extra galactic neutrinos. Even though extra galactic contributions in neutrino flux is subdominant in the energy range for $\nu$BDM~\cite{Vitagliano:2019yzm} , it can lead modification in e.g. spatial and kinetic distributions of $\nu$BDM. All those factors of improvement are reserved for the future work.

%%%%%%%%%%%%%%%%%%%%%%%%%%%%%%%%%%%%%%%%%%%%%%%%%%%%%%%%%%%%%%%%%%%%%%%%%%%%%%%

\section*{Acknowledgments}

The work is supported in part by Basic Science Research Program through the National Research Foundation of Korea (NRF) funded by the Ministry of Education, Science and Technology [NRF-2018R1A4A1025334, NRF-2019R1A2C1089334 (SCP), NRF-2019R1C1C1005073 (JCP) and NRF-2020R1I1A1A01066413 (PYT)].

\bigskip

\bibliographystyle{apsrev4-2}
\bibliography{ref}

\end{document}